\documentclass[prl,showpacs,twocolumn,aps]{revtex4}
\usepackage{rotating}
\usepackage{graphicx}
\newcommand{\be}{\begin{equation}}

\newcommand{\ee}{\end{equation}}
\newcommand{\bea}{\begin{eqnarray}}
\newcommand{\eea}{\end{eqnarray}}

\newcommand{\beq}{\begin{equation}}
\newcommand{\eeq}{\end{equation}}
\newcommand{\beqn}{\begin{eqnarray}}
\newcommand{\eeqn}{\end{eqnarray}}
\newcommand{\bmath}{\begin{subequations}}
\newcommand{\emath}{\end{subequations}}
\begin{document}
\title{Dynamic Hubbard Model: Effect of Finite Boson Frequency}
\author{F. Marsiglio$^{a}$, R. Teshima$^{a}$ and J. E. Hirsch$^{b}$ }
\affiliation{$^{a}$Department of Physics, University of Alberta, Edmonton,
Alberta, Canada T6G 2J1\\
$^{b}$Department of Physics, University of California, San Diego,
La Jolla, CA 92093-0319}

\date{\today}
\begin{abstract}
Dynamic Hubbard models describe coupling of a boson degree of freedom 
to the
on-site electronic double occupancy. In the limit of infinite 
boson frequency this
coupling gives rise to a correlated hopping term 
in the effective Hamiltonian and to
superconductivity when the Fermi 
level is near the top of the band.
Here we study the effect of finite 
boson frequency through a generalized
Lang-Firsov transformation and 
a high frequency expansion. 
It is found that finite frequency 
enhances the tendency to superconductivity in this model.
\end{abstract}
\maketitle
\date{\today}


\section{INTRODUCTION}
\label{sec:int}

An attempt to understand electron-electron interactions in solids over
the last few decades has focused on simplified models which single out, for
example, on-site or nearest neighbour Coulombic interactions. While it is
usually understood that the parameters that are required in these models are
not `bare' parameters, it is often tacitly assumed that the omitted physics
can be recovered by simply modifying these parameters to effectively 
incorporate
the missing physics.  However, the notion that this procedure may in fact leave
out much of the interesting physics has been emphasized by one of us
recently \cite{hirsch01}, and a new class of model Hamiltonians, 
`dynamic Hubbard models' ,
was introduced to remedy this 
difficulty\cite{hirsch01,hirsch02a,hirsch02bc,hirsch03} . These 
Hamiltonians describe
an essential difference between the empty and 
the doubly occupied Wannier
orbital, that the former is representable 
by a single Slater determinant and the
latter one is not. This 
difference is not described by the conventional models
and leads to 
electron-hole asymmetry and to an uncoventional mechanism
of 
superconductivity\cite{hirsch89}.

In this paper we focus attention on one particular dynamic Hubbard model, which
introduces a modulation of the Hubbard $U$ by coupling the electron double
occupancy to a fictitious local boson displacement. The site
Hamiltonian that describes this model is given by \cite{hirsch01}:
\beq
H_i = {p_i^2 \over 2 M} + {1 \over 2} Kq_i^2 + (U + \alpha q_i)n_{i\uparrow}
n_{i\downarrow}
\label{siteh}
\eeq
where the boson is characterized by an Einstein oscillator with mass $M$ and
spring constant $K$. The displacement of the local oscillator $q_i$ is coupled
to the electron double occupancy with coupling constant $\alpha$, and 
as a consequence
the on-site repulsion $U + \alpha q_i$ becomes a 
dynamical variable . The 
Hamiltonian in electron representation is 
given by
\beqn
H = \omega_0 \sum_i a_i^\dagger a_i - t \sum_{i \delta \atop \sigma}
\biggl(
c_{i\sigma}^\dagger c_{i + \delta \sigma} + c_{i+\delta\sigma}^\dagger 
c_{i\sigma}
\biggr) - \mu \sum_{i\sigma} n_{i\sigma}
\nonumber \\
+\sum_i \biggl[U + g\omega_0(a_i + 
a_i^\dagger)\biggr]n_{i\uparrow}n_{i\downarrow}, \ \ \ \ \ \ \ \ \ \
\label{electronh}
\eeqn
where $c_{i\sigma}^\dagger$ is an electron creation operator 
for site $i$ and spin
$\sigma$, $n_{i\sigma}$ is the corresponding  number operator, and we have
introduced the boson creation and annihilation operators, $a_i^\dagger$ and
$a_i$, respectively. The Einstein oscillator frequency is $\omega_0 \equiv
(K/M)^{1/2}$, and $g \equiv \alpha/(2K\omega_0)^{1/2}$. In addition we
have introduced the electron kinetic energy, with nearest neighbour hopping
amplitude $t$ and chemical potential $\mu$. Finally, $U$ describes the
static electron-electron repulsion.

Treating the four-fermion term 
involving the boson in mean field leads to a model
with ordinary 
Holstein-like electron-boson 
coupling
\beq
H_{el-b}=g(n)\omega_0(a_i^\dagger+a_i)(n_{i\uparrow}+n_{i\downarrow})
\eeq
\beq
g(n)=\frac{n}{2}g
\eeq
where 
the coupling constant $g(n)$ increases with electron occupation and 
is maximum
for the Fermi level at the top of the band. Hence the 
quasiparticle dressing in this
model increases as the Fermi level 
rises in the band, and conversely it decreases as the system
is doped 
with holes. There are indications in the high $T_c$ cuprates that 
the
quasiparticle dressing decreases with hole 
doping\cite{experiments}
 and this is one of the motivations to study 
this Hamiltonian. However to fully understand
the physics of this 
model the mean field decoupling leading to Eqs. (3),(4) is 
certainly 
inappropriate.

We study the Hamiltonian Eq. (2) in hole rather than electron 
representation.
A particle-hole
transformation yields in addition a Holstein-like coupling:
\beqn
H = \omega_0 \sum_i a_i^\dagger a_i - t \sum_{i \delta \atop \sigma}
\biggl(
c_{i\sigma}^\dagger c_{i + \delta \sigma} + 
c_{i+\delta\sigma}^\dagger c_{i\sigma}
\biggr) - \mu \sum_{i\sigma} n_{i\sigma}
\nonumber \\
-g\omega_0\sum_{i\sigma}(a_i + a_i^\dagger) n_{i\sigma}
+\sum_i \biggl[U + g\omega_0(a_i + 
a_i^\dagger)\biggr]n_{i\uparrow}n_{i\downarrow},
\label{holeh}
\eeqn
where now $c_{i\sigma}^\dagger$, $n_{i\sigma}$  are hole creation 
operator and hole number operator respectively.  
One of the 
essential ingredients of this and other similar models is that
the system is inherently {\it not} electron-hole symmetric. This is clear
within the Hartree approximation, but the asymmetry is best revealed through
a generalized Lang-Firsov transformation \cite{hirsch00ab,hirsch01}.

In the following section we perform this transformation, and derive a
Migdal-like expansion valid in the anti-adiabatic limit, which allows 
us to study
corrections due to finite boson frequency. While the 
antiadiabatic limit
is very well understood\cite{hirsch89}, the effect of non-infinite 
boson frequency (retardation)
is not . The purpose of this paper is to address this question which 
is of fundamental
importance because derivation of the model from 
first principles  necessarily leads to finite boson 
frequency\cite{hirsch02a}.  The expansion is performed to first order 
in inverse
boson frequency, and the impact on superconductivity is determined. 
We find that
retardation increases $T_c$ at all hole densities, and increases the range of
hole densities over which superconductivity occurs. This is in 
qualitative agreement
with a previous study \cite{hirsch02bc} which determined the effective pairing
interaction for two holes within a pseudospin model on small 
clusters. That study
found an increased attractive interaction due to retardation as well.

\section{GENERALIZED LANG-FIRSOV TRANSFORMATION}
\label{sec:gen}

We use the transformation
\be
\tilde{H} = e^G H e^{-G},
\label{lf_trans}
\ee
where \cite{hirsch00ab}
\be
G = \sum_i g\bigl(a_i - a_i^\dagger\bigr)\bigl(n_{i\uparrow} + 
n_{i\downarrow} -
n_{i\uparrow} n_{i\downarrow}\bigr).
\label{lf_g}
\ee
Application of Eq. (\ref{lf_trans}) results in
\bea
\tilde{H} = \omega_0 \sum_i a_i^\dagger a_i - \mu_0 \sum_{i\sigma} n_{i\sigma}
+ U_{\rm eff} \sum_i n_{i\uparrow}n_{i\downarrow}
\nonumber \\
-t \sum_{i \delta \atop \sigma} \biggl( X_{i\sigma}^\dagger X_{i+\delta \sigma}
c_{i\sigma}^\dagger c_{i + \delta \sigma} + H.c. \biggr),
\label{trans_h}
\eea
where $U_{\rm eff} \equiv U - \omega_0g^2$, $\mu_0 \equiv \mu + \omega_0g^2$,
and
\be
X_{i\sigma}^\dagger \equiv \exp{\biggl(g(a_i - a_i^\dagger)(1 - n_{i-\sigma})
\biggr)},
\label{xop}
\ee
dresses the hole-hopping amplitudes. These operators are dependent on both
the oscillator and hole degrees of freedom.

Eq. (\ref{trans_h}) leads to a low energy Hamiltonian for the hole degrees of
freedom if the ground state expectation value is taken with respect to
the oscillator degrees of freedom. Following Ref. \cite{hirsch01}, we find
\bea
\tilde{H} \approx -\tilde{t_0} \sum_{i \delta \atop \sigma} \biggl(
c_{i\sigma}^\dagger c_{i + \delta \sigma} + H.c. \biggr) -
\mu_0 \sum_{i\sigma} n_{i\sigma} \ \ \ \ \ \ \ \
\nonumber \\
+ U_{\rm eff} \sum_i n_{i\uparrow}n_{i\downarrow}
-\Delta t \sum_{i \delta \atop \sigma} \biggl(
c_{i\sigma}^\dagger c_{i + \delta \sigma} + H.c. \biggr)
\biggl( n_{i-\sigma} + n_{i+\delta -\sigma} \biggr)
\label{antiad}
\eea
where
\be
\tilde{t_0} \equiv te^{-g^2}
\label{ttilde}
\ee
and
\be
\Delta t  \equiv te^{-g^2} \bigl( e^{g^2/2} - 1 \bigr).
\label{deltat}
\ee
Eq. (\ref{antiad}) is the effective quasiparticle Hamiltonian studied in Ref.
\cite{hirsch89}. This model is known to lead to superconductivity at low hole
dopings, and to characteristic properties related to `undressing' phenomenology
\cite{hirsch00ab}. The effective single hole 
hopping
\be
\tilde{t}(n)=\tilde{t}_0+n\Delta t
\ee
is an increasing 
function of hole doping, and so is the effective hole
bandwidth.

Based on previous studies\cite{hirsch89}, superconductivity (and in 
fact pair binding)
occurs in the model Eq. (\ref{trans_h}) in the anti-adiabatic limit. We wish
to answer the question: will superconductivity occur more or less readily
with retardation, i.e. away from the anti-adiabatic limit~? To this end, we
use perturbation theory following the generalized Lang-Firsov transformation.
That is we note that the operator $X_{i\sigma}^\dagger$ can be written
\bea
X_{i\sigma}^\dagger & = & n_{i-\sigma} + (1-n_{i-\sigma})\exp{\bigl( g(a_i
-a_i^\dagger) \bigr)}
\nonumber \\
& = & n_{i-\sigma} + (1-n_{i-\sigma})e^{-g^2/2} e^{-ga_i^\dagger}e^{ga_i}
\nonumber \\
& \approx & n_{i-\sigma} + (1-n_{i-\sigma})e^{-g^2/2} \bigl(1 +
g(a_i - a_i^\dagger) \bigr),
\label{expansion1}
\eea
and similarly for $X_{i\sigma}$. In the last line of Eq. (\ref{expansion1})
we have expanded the exponential to first order in the oscillator momentum
(or, equivalently, the displacement). To put terms involving the linear
oscillator operators into `standard' form (with the oscillator displacement
rather than the momentum), we utilize the canonical transformation
\bea
a_i & \rightarrow & -i a_i
\nonumber \\
a_i^\dagger & \rightarrow & i a_i^\dagger.
\label{canon}
\eea
Finally, Fourier transforming all the operators to momentum space, we obtain
\bea
\tilde{H} &=& \omega_0 \sum_q a_q^\dagger a_q + \sum_{k\sigma} \bigl(
\tilde{\epsilon_k} - \tilde{\mu_0} \bigr) c_{k\sigma}^\dagger c_{k\sigma}
\nonumber \\
& & + {1 \over N} \sum_{k k^\prime} \biggl[ U_{\rm eff} + 2{\Delta t \over
\tilde{t_0} + n\Delta t}(\tilde{\epsilon_k} + 
\tilde{\epsilon_k^\prime}) \biggr]
c_{k\uparrow}^\dagger c_{-k\downarrow}^\dagger c_{-k^\prime \downarrow}
c_{k^\prime \uparrow}
\nonumber \\
& & + {1 \over \sqrt{N}}\sum_{k k^\prime \atop \sigma} g_{kk^\prime}
\bigl(a_{k-k^\prime}+a_{-(k-k^\prime)}^\dagger \bigr)
c_{k\sigma}^\dagger c_{k^\prime \sigma},
\label{ham_k}
\eea
where
\bea
\tilde{\epsilon_k} &=& -2(\tilde{t_0} + n\Delta t)\sum_\delta \cos{(k\delta)},
\label{def1}
\\
\tilde{t_0} &=& te^{-g^2},
\label{def2}
\\
\Delta t &=& \tilde{t_0}(e^{g^2/2} - 1),
\label{def3}
\\
U_{\rm eff} &=& U - \omega_0 g^2
\label{def4}
\\
\tilde{\mu_0} &=& \mu + \omega_0 g^2 - U_{\rm eff}n/2,
\label{def5}
\eea
and
\be
g_{kk^\prime} \equiv ig {\tilde{t_0}(1 - n/2) + n\Delta t/2 \over
\tilde{t_0} + n\Delta t} \bigl( \tilde{\epsilon_k} - 
\tilde{\epsilon_{k^\prime}}
\bigr).
\label{def6}
\ee
Note that we have included Hartree corrections already in the hopping and
chemical potential terms; these are not to be included again when treating the
Hamiltonian (\ref{ham_k}). The correction to the hole hopping matrix element
in particular leads to a wider bandwidth $\tilde{D}$ for increasing 
number of
holes in the band.

\section{ELIASHBERG-LIKE APPROXIMATION FOR HIGH BOSON FREQUENCIES}
\label{sec:eli}

Eq. (\ref{ham_k}) looks like a standard Hamiltonian with linear electron-phonon
coupling. One can determine its properties with the Migdal-Eliashberg
approximation. Omitting the detailed steps in the derivation, the final
result is:
\bea
& & \phi(k,i\omega_m) = 
\nonumber \\
& & {1 \over N\beta}\sum_{k^\prime,m^\prime}\biggl(
{2\omega_0 g_{-k^\prime -k} g_{k^\prime k} \over \omega_0^2 +
(\omega_m - \omega_{m^\prime})^2 } - V_{kk^\prime}
\biggr)
{\phi(k^\prime,i\omega_{m^\prime}) \over E(k^\prime,i\omega_{m^\prime})
} \ \ \ \ \ \ \ \
\label{elias1} \\
& & \Sigma(k,i\omega_m) = 
\nonumber \\
& & {1 \over N\beta}\sum_{k^\prime,m^\prime}\biggl(
{2\omega_0 g_{k^\prime k} g_{k k^\prime} \over \omega_0^2 +
(\omega_m - \omega_{m^\prime})^2 }
\biggr)
{\tilde{G_0}^{-1}(-k^\prime,-i\omega_{m^\prime}) \over 
E(k^\prime,i\omega_{m^\prime})
} \ \ \ \ \ \ \ \
\label{elias2}
\eea
where $\phi(k,i\omega_m)$ and $\Sigma(k,i\omega_m)$ are the pairing and normal
self energies, and the denominator $E(k,i\omega_m)$ is given by
\be
E(k,i\omega_m) \equiv \tilde{G_0}^{-1}(k,i\omega_{m})
\tilde{G_0}^{-1}(-k,-i\omega_{m})+\phi^2(k,i\omega_{m})
\label{denom}
\ee
with
\be
\tilde{G_0}^{-1}(k,i\omega_m) \equiv i\omega_m - (\tilde{\epsilon_k} -
\tilde{\mu_0}) -\Sigma(k,i\omega_m).
\label{g0}
\ee
These equations are standard Eliashberg equations, with
$i\omega_m \equiv i\pi T (2m - 1)$ a fermion Matsubara frequency, $\beta
\equiv 1/(k_BT)$ the inverse temperature, and $N$ the number of lattice sites.
The direct electron-electron interaction is given by
\be
V_{kk^\prime} = U_{\rm eff} + {2 \Delta t \over \tilde{t_0} + n \Delta t}
\bigl(\tilde{\epsilon_k} + \tilde{\epsilon_{k^\prime}} \bigr).
\label{vcoul}
\ee
It is also customary to write the normal self energy in terms of an odd and
even (in Matsubara frequency) part:
\be
\Sigma(k,i\omega_m) \equiv i\omega_m(1 - Z(k,i\omega_m)) + \chi(k,i\omega_m),
\label{even_odd}
\ee
which leads to three coupled equations instead of the two Eqs.
(\ref{elias1},\ref{elias2}). Note that the arguments of the g-factors
in the kernels of Eqs. (\ref{elias1},\ref{elias2}) are actually different,
and lead, in this case, to kernels with opposite signs in the pairing and
normal equations. These equations also require an auxiliary number equation:
\bea
& & n = 1 + 2 {\rm Re} {1 \over N\beta} 
\nonumber \\
& & \sum_{k,m}
{\tilde{G_0}^{-1}(-k,-i\omega_{m}) \over
\tilde{G_0}^{-1}(k,i\omega_{m})
\tilde{G_0}^{-1}(-k,-i\omega_{m})+\phi^2(k,i\omega_{m})
}, \ \ \ \ \ \
\label{number}
\eea
where $n$ is the hole number density.

These equations are most easily solved by noting that each unknown function
can be decomposed into three k-dependent pieces. For example,
\be
\phi(k,i\omega_m) \equiv
\phi_0(i\omega_m) +
\phi_1(i\omega_m)\biggl( -{\tilde{\epsilon_k} \over \tilde{D}/2}\biggr) +
\phi_2(i\omega_m)\biggl( {\tilde{\epsilon_k} \over \tilde{D}/2} \biggr)^2
\label{decompose}
\ee
where $\tilde{D} \equiv \tilde{D}(n) = 8(\tilde{t_0} + n\Delta t)$
is the $n$-dependent hole bandwidth. This makes the
entire problem only slightly more difficult than the standard Eliashberg
equations which require iterative solution in Matsubara frequency space
\cite{marsiglio03}.

Nonetheless, we need not solve these rather complicated equations to determine
their properties for large boson frequency, $\omega_0$. Eqs.
(\ref{elias1},\ref{elias2}) both contain a kernel of the form:
\bea
\kappa_{k k^\prime}(i\nu_n) & = & - {\tilde{D}(n) \over \omega_0} {\omega_0^2
\over \omega_0^2 + \nu_n^2}{g^2 \over 2} \biggl(
{\tilde{t_0}(1 - n/2) + n\Delta t/2 \over
\tilde{t_0} + n\Delta t}
\biggr)^2
\nonumber \\
& & \biggl( {\tilde{\epsilon_k} - \tilde{\epsilon_{k^\prime}} \over
\tilde{D}(n)/2 } \biggr)^2,
\label{kernel}
\eea
where $i\nu_n \equiv i2\pi Tn$ is a Boson Matsubara frequency, and the
other variables have been previously defined. We have written
Eq.~(\ref{kernel}) with an explicit prefactor $\tilde{D}(n)/ \omega_0$,
which shows that for infinite boson frequency, $\omega_0$, all of the
complications due to the electron-boson coupling can be ignored. This leaves
\bea
& & Z(k,i\omega_m) = 1,
\nonumber \\
& & \chi(k,i\omega_m) = 0,
\nonumber \\
& &\phi(k,i\omega_m) \equiv \phi(k)
\nonumber \\
& & = -{1 \over N\beta} \sum_{k^\prime,m^\prime}
V_{kk^\prime}
{\phi(k^\prime) \over
\tilde{G_0}^{-1}(k^\prime,i\omega_{m^\prime})
\tilde{G_0}^{-1}(-k^\prime,-i\omega_{m^\prime})+\phi^2(k^\prime) }, \ \ \ \ \ \ \ \ \ \ \
\label{bcslike}
\eea
which is a straightforward BCS-like problem which has been solved previously
\cite{hirsch89}. The result is an extended s-wave solution for
superconductivity at low hole densities. To understand what happens for
large but non-infinite boson frequencies, we adopt a standard approximation
for Eliashberg theory, i.e. in the kernel we assume that the boson frequency
is much larger than the other energy scales, so that it becomes independent of
Matsubara frequency. This allows us to neglect the normal channels, and
focus on the gap equation \cite{note1}:
\bea
& & \phi(k) = {1 \over N\beta} \tilde{D}(n) 
\nonumber \\
& & \sum_{k^\prime,m^\prime} \biggl[
-\kappa(n) \bigl({ \tilde{\epsilon_k} - \tilde{\epsilon_{k^\prime}}
\over
\tilde{D}(n)/2 } \bigr)^2
- {V_{k k^\prime} \over \tilde{D}(n) }
\biggr]
{\phi(k^\prime) \over \omega_{m^\prime}^2 + (\tilde{\epsilon_{k^\prime}}
-\tilde{\mu_0})^2}, \ \ \ \ \ \ \ \ \
\label{bcs_renorm}
\eea
where
\be
\kappa(n) \equiv {\tilde{D}(n) \over \omega_0}
{g^2 \over 2} \biggl(
{\tilde{t_0}(1 - n/2) + n\Delta t/2 \over
\tilde{t_0} + n\Delta t}
\biggr)^2.
\label{kernel_n}
\ee
The Matsubara sum in this equation is readily performed:
\bea
& &{1 \over \beta} \sum_m {\tilde{D}(n)/2 \over \omega_m^2 +
(\tilde{\epsilon}_k - \tilde{\mu}_0)^2 } =
{ \tilde{D}(n)/2 \over 2 (\tilde{\epsilon}_k - \tilde{\mu}_0)} \bigl( 1 - 2
f(\tilde{\epsilon}_k - \tilde{\mu}_0) \bigr)
\nonumber \\
& &\equiv F_1(\tilde{\epsilon}_k).
\label{mats}
\eea
As noted already in Eq. (\ref{decompose}), the $k$-dependence can
be decomposed into a constant, a term proportional to $\tilde{\epsilon}_k$,
and a third term proportional to $\tilde{\epsilon}_k^2$. The latter
term is new, and arises from the fact that a non-infinite boson frequency
is used. More explicitly, if we use a constant electron density of states
($g(\tilde{\epsilon}) = 1/\tilde{D}, \ \ -\tilde{D}/2 < \tilde{\epsilon}
< \tilde{D}/2$), then we have:
\bea
& & \phi(x) = \int_{-1}^1 \ dx^\prime \  \phi(x^\prime) F_1(\tilde{\epsilon}^\prime)
\nonumber \\
& & \biggl( - \kappa(n) \bigl(
x^2 + x^{\prime 2} \bigr) + 2 \kappa(n) x x^\prime - k(x + x^\prime)
-u_{\rm eff} \biggr),
\ \ \ \ \
\label{bcseqn}
\eea
where $u_{\rm eff} \equiv U_{\rm eff}/\tilde{D}(n)$, $k \equiv
8 \Delta t /\tilde{D}(n)$, and $x \equiv \tilde{\epsilon}/ (\tilde{D}(n)/2)$,
and similarly for the primed quantities. Normally, positive components of the
kernels contribute to pairing, and hence to $T_c$. For example, focusing
on the anti-adiabatic terms, $u_{\rm eff}$ clearly deters pairing, whereas
$k$ will enhance pairing in the energy range where holes dominate (negative
$\tilde{\epsilon}$ and negative $\tilde{\epsilon}^\prime$). At the same time
$k$ hurts pairing in the electron regime (positive $\tilde{\epsilon}$).
Clearly the effect of $\kappa(n)$ is helpful (to pairing) in the central
term (and hence adds to the role of $k$), and is detrimental in the
first term. However, a full solution is required to
determine the overall impact of a non-zero $\kappa(n)$ (and hence non-infinite
$\omega_0$ --- see Eq. (\ref{kernel_n})) \cite{note2}.

\section{BCS-LIKE SOLUTION}
\label{sec:bcs}

A full solution to Eq. \ref{bcseqn} is obtained by noting that
\be
\phi(x) = a_0 - a_1 x + a_2x^2.
\label{phi}
\ee
Substitution of Eq. \ref{phi} into Eq. \ref{bcseqn} leads to 3 homogeneous
eqns, so that $T_c$ is given by setting the determinant of
the following 3X3 matrix to
zero:
\be
\pmatrix{ 1 + \kappa(n)I_2 + u I_0 -kI_1 &
\kappa(n) I_3 + uI_1 - kI_2 &
\kappa(n) I_4 + uI_2 - k I_3 \cr
-kI_0 - 2 \kappa(n) I_1 &
1 - kI_1 - 2 \kappa(n) I_2 &
-kI_2 - 2 \kappa(n) I_3 \cr
\kappa(n) I_0 & \kappa(n) I_1 & 1 + \kappa(n) I_2 \cr } = 0,
\label{det}
\ee
where
\be
I_\ell \equiv \int_{-\tilde{D}(n)/2}^{\tilde{D}(n)/2} \ \ d
\tilde{\epsilon} \biggl(
-{\tilde{\epsilon} \over \tilde{D}(n)/2} \biggr)^\ell { 1 - 2 
f(\tilde{\epsilon}
-\tilde{\mu}_0) \over 2 (\tilde{\epsilon}-\tilde{\mu}_0) }.
\label{i_int}
\ee
For the number equation we assume all non-pairing interactions are already
taken into account in the parameters; thus we obtain, for the chemical
potential:
\be
{\tilde{\mu}_0 \over \tilde{D}(n)/2} = -(1 - n),
\label{mu}
\ee
where $n$ is the hole density. The evaluation of the determinant, and
the determination of $T_c(n)$ is a tedious process; in weak coupling,
one can use the relations:
\bea
I_0 &=& \ln{\biggl({1.13 \over k_BT_c} {\tilde{D}(n) \over 2} \sqrt{n(2-n)}
\biggr)},
\nonumber \\
I_1 &=& \rho (I_0 - 1)
\nonumber \\
I_2 &=& \rho^2 I_0 + {1 \over 2} - {3 \over 2} \rho^2
\nonumber \\
I_3 &=& \rho^3 I_0 + {1 \over 2}\rho - {11 \over 6} \rho^3
\nonumber \\
I_4 &=& \rho^4 I_0 + {1 \over 4} + {1 \over 2}\rho^2 - {25 \over 12} \rho^4,
\label{identities}
\eea
where $\rho \equiv 1 - n$. Weak coupling is accurate over a wide range of
parameters \cite{marsiglio90}. The end result is
\be
k_BT_c = 1.13 {\tilde{D}(n) \over 2} \sqrt{n(2-n)}\exp{(-a/b)},
\label{tc}
\ee
where
\bea
a &=&
1 + k\rho (2 + k\rho) - \kappa \rho (-k - 2u\rho + 3k\rho^2) +
\nonumber \\
& & \kappa^2(-9 + 30\rho^2 + 7\rho^4)/12 -
\kappa^3(3 - 9\rho^2 + 13\rho^4 + \rho^6)/12
\nonumber \\
b &=&
(k^2 - 2u + 4k\rho + k^2\rho^2)/2 +
\kappa(3u + 3u\rho^2 - 4k\rho^3)/3 +
\nonumber \\
& &
\kappa^2(1 + 6\rho^2 + \rho^4)/4 -
\kappa^3(9-9\rho^2+15\rho^4+\rho^6)/36,
\label{a_and_b}
\eea
with $\kappa \equiv \kappa(n)$.
Note that if only the coupling to the boson were present, then, for
small hole density, $b/a \approx 2\kappa^2$ for $\kappa << 1$. Thus
the boson leads to an attractive interaction, independent of the sign
of $\kappa$, but $T_c$ is unobservably small if $\kappa <<1$. However,
if $T_c$ already exists (through the presence of the anti-adiabatic terms,
in this case), then the enhancement due to the coupling to the boson (which,
in this case, means a coupling at non-infinite frequency) can be substantial.
This nonlinear effect is merely due to the nature of the exponential function.

In Fig. 1 we plot $T_c$ vs. hole density $n$ for parameter values
$\Delta t=0.185$ eV, $U_{\rm eff} = 5.0$ eV, $\tilde{t}_0 = 0.025$ eV,
with $\omega_0 = \infty, 10$ eV, and $5$ eV.
The explicit $n$ dependence of $\kappa(n)$ is not significant, and is shown
in Fig. 2, for a variety of parameter values (note: for Fig. 1,
$\Delta t/\tilde{t}_0 = 7.4$).
Fig. 1 shows that $T_c$ is enhanced by retardation --- as $\omega_0$ decreases, 
$\kappa_0 \equiv \kappa(n = 0)$ increases. It is also clear that the range 
of hole densities increases with increasing amount of retardation.

Finally, we can compute the effective mass and quasiparticle residue as a function
of doping. These are obtained in a two-step process. First, in the anti-adiabatic
limit ($\omega_0 \rightarrow \infty$), the effective mass ratio is given for low
hole densities \cite{hirsch00ab} as
\be
m_{\rm aa}^\ast/m = {(1 + \Delta t/\tilde{t}_0)^2 \over (1 + n \Delta t/\tilde{t}_0)},
\label{effm_inf}
\ee
and the  quasiparticle residue is similarly given by \cite{hirsch00ab}
\be
z_{{\rm aa} 0}={(1 + {n \over 2}\Delta t/\tilde{t}_0)^2 \over (1 + \Delta t/\tilde{t}_0)^2}.
\label{z0_inf}
\ee
Here $z_{{\rm aa} 0}$ is the quasiparticle residue on the Fermi surface, and
the subscript $aa$ stands for the anti-adiabatic limit. To incorporate 
changes due to retardation effects, we proceed 
in the usual fashion, and compute the real part of the analytic continuation to 
the real axis of the self energy expression
Eq. (\ref{elias2}) in the normal state, and without self-consistency (i.e. we
omit the self energy on the right hand side). Then
\be
{m^\ast \over m} = {m_{\rm aa}^\ast \over m}\biggl( {1 - a_\omega \over 1 + a_k} \biggr)
\label{effm}
\ee
and
\be
z_0= {z_{{\rm aa} 0} \over 1 - a_\omega},
\label{z0}
\ee
where 
\be
a_\omega \equiv { \partial \Sigma_1 \over \partial \omega}(\tilde{\epsilon_k}, \omega
+i\delta) |_{\omega = 0, \tilde{\epsilon_k} = \tilde{\mu}_0}
\label{a_w}
\ee
and
\be
a_k \equiv { \partial \Sigma_1 \over \partial \tilde{\epsilon_k}}(\tilde{\epsilon_k}, \omega
+i\delta) |_{\omega = 0, \tilde{\epsilon_k} = \tilde{\mu}_0},
\label{a_k}
\ee
where $\Sigma_1$ is the real part of the self energy.
These quantities are given by
\be
a_\omega = 2\kappa(n) \biggl(1 + (1-n)^2 \biggr)
\label{a_w_result}
\ee
and 
\be
a_k = \kappa(n) \biggl(1 + (1-n)^2 \biggr).
\label{a_k_result}
\ee
The effective mass and residue are plotted in Fig. 3. It is clear that the
effective mass decreases and the quasiparticle residue increases as a function
of increasing hole concentration, as expected in this model. Note that
retardation actually {\it decreases} the effective mass and {\it increases} the
quasiparticle residue, because retardation `undoes' some of the effects from
the generalized Lang-Firsov transformation. Note that retardation has little
effect on the doping dependence of these properties. To emphasize the observation
made in Ref. \cite{hirsch02bc} we have plotted the inverse effective mass ratio
in Fig. 3b (dashed curve) in the anti-adiabatic limit, to show that the
residue and inverse mass behave similarly as a function of doping. This indicates
that the frequency dependence of the hole self energy is most important (as opposed to
the momentum dependence).

\section{CONCLUSIONS}
We have studied the effect of finite boson
frequency in a dynamic Hubbard model. The model describes 
modulation of the Hubbard on-site repulsion by a   boson
degree of freedom. In the limit of infinite boson frequency the properties
of the model are well understood, and here we studied the effect of 
finite frequencies through a generalized Lang-Firsov transformation 
and a high frequency expansion.

One might also ask about the use of cruder approximations. For example, one
might first consider the Hamiltonian (\ref{holeh}) in the mean field 
decoupling given by Eqs. (3,4). Doing then the usual Lang-Firsov 
transformation would yield an effective repulsive Hubbard model, with no superconductivity
in the anti-adiabatic limit at the BCS-Eliashberg level of approximation.
Thus the starting point for such a  study would already be in serious error.
The dynamics inherent in the coupling of the double occupancy of holes to
the boson displacement given in Eq. (\ref{holeh}) is crucial for the
occurrence of pairing.

The central result of this work is the finding that 
$T_c$ increases with retardation .
That this effect is expected was also indicated by the exact diagonalization
study in Ref. \cite{hirsch02bc}.  In the derivation of this model 
from first principles\cite{hirsch01,hirsch02a} the boson frequency 
represents the scale of intra-atomic electronic excitation energies. 
This scale is expected to be large compared to the scale of 
interatomic hopping, but is certainly not infinite. Therefore it is 
essential to study the effect of finite frequency corrections as done 
in this paper.

This finding supports the possibility that the model may be 
relevant to the superconductivity of real materials. It should also 
be noted that in the microscopic derivation of the model both the 
coupling constant $g$ increases and the boson frequency decreases as 
the degree of negative charging of the ion increases. According to 
the results of this study, both effects enhance superconductivity
in this model. Hence our results support the hypothesis that conduction 
of holes through negative ions is conducive to high temperature 
superconductivity.  This is  in qualitative agreement with the fact 
that high $T_c$ is found in cuprates (with holes in $O^=$ ions) and 
in $MgB_2$ (with holes in $B^- $ ions).

\acknowledgments
We are grateful to the Natural Sciences and
Engineering Research Council (NSERC) of Canada and the Canadian
Institute for Advanced Research (CIAR) for support.

\bibliographystyle{prb}

\begin{thebibliography}{1}
\bibitem{note1} Strictly speaking, $\chi(k,i\omega_m)$ simply renormalizes
the chemical potential, but $Z(k,i\omega_m)$ remains dependent on Matsubara
frequency, even if $\kappa_{k k^\prime}(i\nu_n)$ is replaced by a constant
independent of Matsubara frequency. These corrections have been neglected,
and are expected to cause only minor quantitative changes.
\bibitem{note2} This is apparent by inspection of the kernel as written
in Eq. \protect{\ref{bcs_renorm}}, where it {\it would appear} that the
additional electron-boson coupling could only have a detrimental effect
on $T_c$ (because of the minus sign in front of the $\kappa(n)$ term).
That this is not the case is clear from the numerical solutions (see Fig. 1).
\end{thebibliography}


\begin{figure}[tp]
\begin{center}
\includegraphics[height=2.0in,width=3.0in]{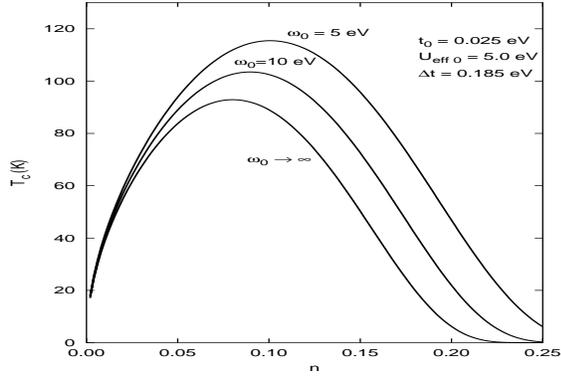}
\caption{
$T_c$ vs. hole doping, $n$. For these parameters $g=2.063$, the hole 
bandwidth at the top of the band is 
$\tilde{D}=0.2$ eV and the electron 
bandwidth at the bottom of the band
is $D=14$ eV. The quasiparticle 
weight is $1$ at the bottom of the band (electrons)
and $0.014$ at 
the top of the band (holes). The boson frequencies are 
$\omega_0=10$ and $5$ eV, which gives
$\kappa_0=0.043$ and $\kappa_0=0.085$ 
respectively.
}
\end{center}
\end{figure}

\begin{figure}[tp]
\begin{center}
\includegraphics[height=2.0in,width=3.0in]{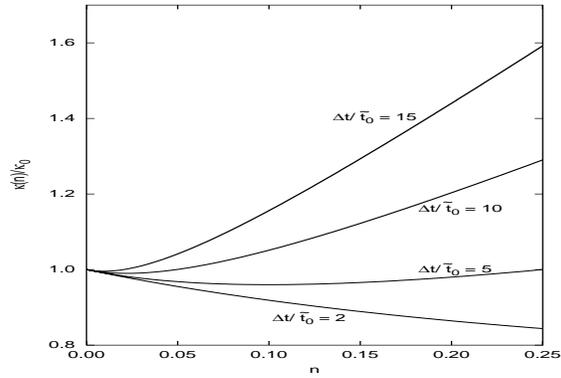}
\caption{
$\kappa(n)$ vs. hole doping, $n$, from Eq. (\protect\ref{kernel_n}).
}
\end{center}
\end{figure}

\begin{figure}[tp]
\begin{center}
\includegraphics[height=2.0in,width=3.0in]{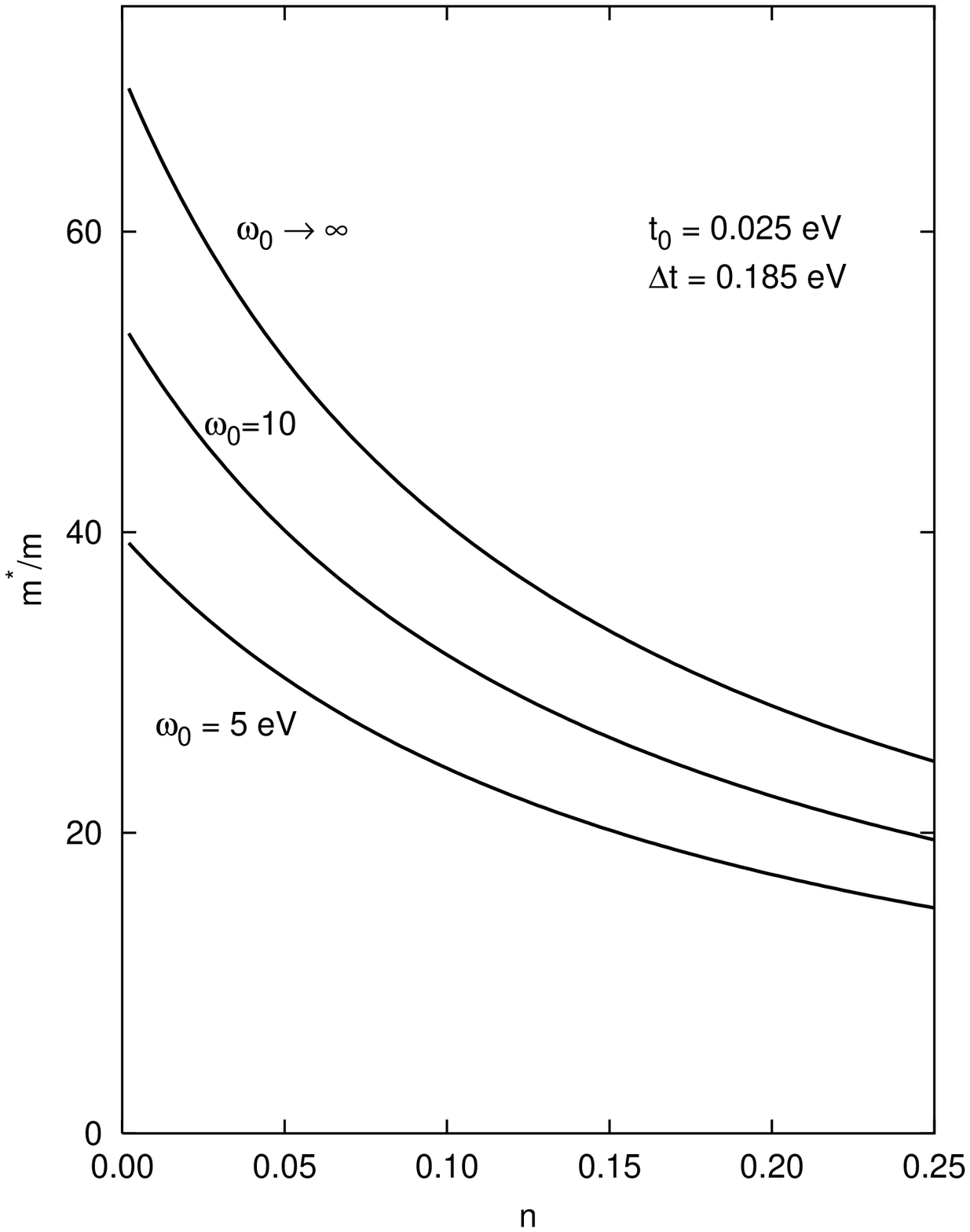}
\includegraphics[height=2.0in,width=3.0in]{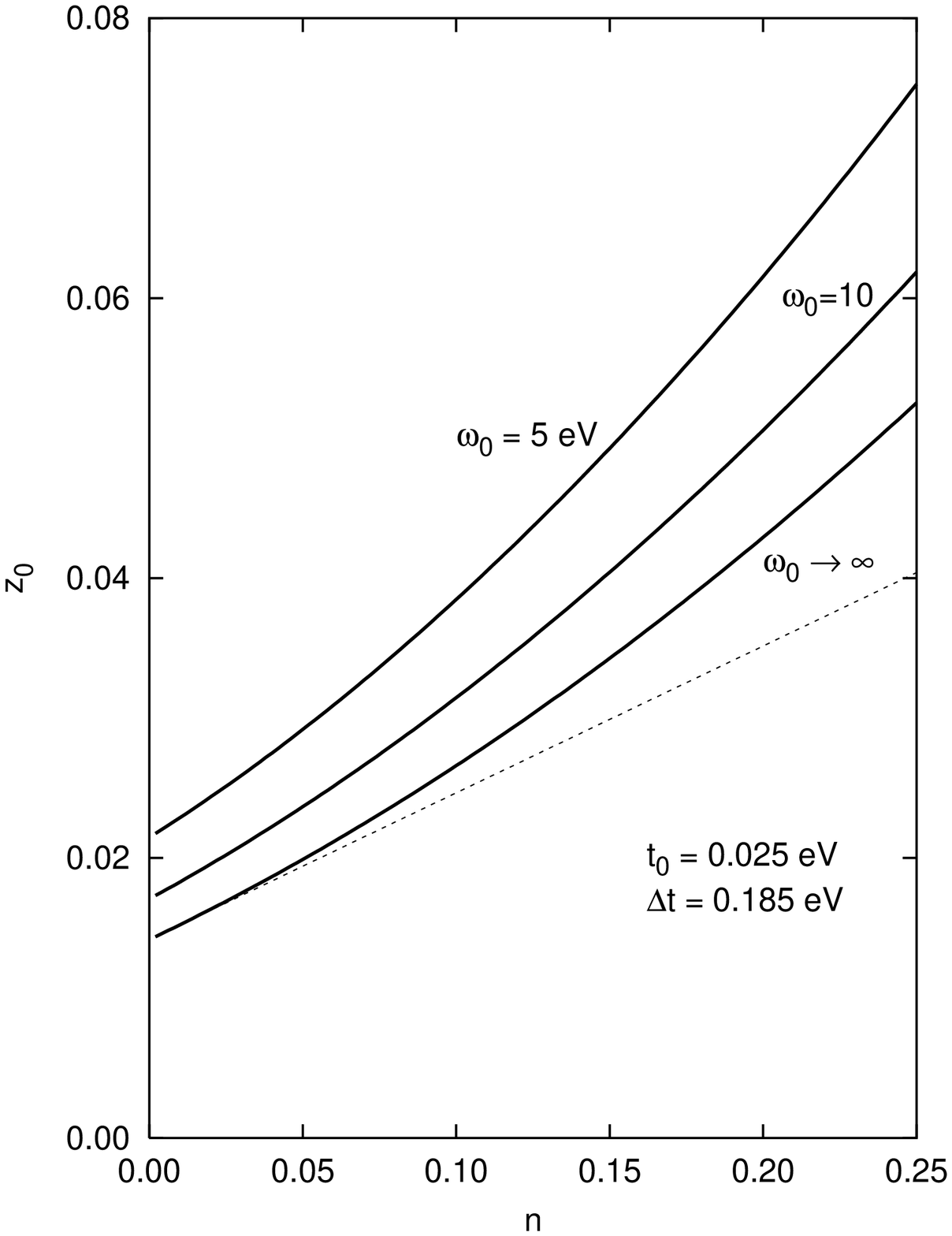}
\caption{
Effective mass (a) and quasiparticle residue (b) vs. hole doping, $n$, for several boson
frequencies as indicated. As expected, the quasiparticle effective mass decreases as
the situation becomes more electron-like; similarly the residue increases towards unity.
Note that both quantities have the same tendency, as is made clear by the dashed curve
in (b), where the inverse effective mass is plotted for $\omega_0 \rightarrow \infty$.
}
\end{center}
\end{figure}

\end{document}